\documentstyle[12pt,epsfig]{article}

\newcommand\M[2]{|{\cal{M}}^{#1}_{#2}|^2}
\newcommand\as{\alpha_{\mathrm{S}}}
\newcommand\smfrac[2]{{\textstyle\frac{#1}{#2}}}
\def\hf{\smfrac{1}{2}}

\def\ep{\epsilon}
\def\ee{$e^+e^-$}
\def\beq{\begin{equation}}
\def\eeq{\end{equation}}
\def\beeq{\begin{eqnarray}}
\def\eeeq{\end{eqnarray}}
\def\cm{{\cal M}}
\def\ket#1{|{#1}>}
\def\bra#1{<{#1}|}
\def\mket#1{|{#1}>_m}
\def\mbra#1{{}_m\!\!<{#1}|}
\def\oket#1{|{#1}>_{m+1}}
\def\obra#1{_{m+1}\!<{#1}|}

\def\bom#1{{\mbox{\boldmath $#1$}}}
\def\to{\rightarrow}

\def\np#1#2#3{Nucl.\ Phys.\ B#1 (19#3) #2}

\def\pr#1#2#3{Phys.\ Rev.\ D #1 (19#3) #2}
\def\prep#1#2#3{Phys.\ Rep.\ #1 (19#3) #2}
\def\prl#1#2#3{Phys.\ Rev.\ Lett.\ #1 (19#3) #2}

\def\zp#1#2#3{Zeit.\ Phys.\ C#1 (19#3) #2}

\begin{document}

\begin{titlepage}
\renewcommand{\thefootnote}{\fnsymbol{footnote}}
\begin{flushright}
     CERN-TH/96-28 \\ hep-ph/9602277
     \end{flushright}
\par \vspace{10mm}
\begin{center}
{\Large \bf
The Dipole Formalism for the Calculation \\[1ex]
of QCD Jet Cross Sections at \\[1ex]
Next-to-Leading Order\footnote{Research supported in part by
EEC Programme {\it Human Capital and Mobility}, Network {\it Physics at High
Energy Colliders}, contract CHRX-CT93-0357 (DG 12 COMA).}}
\end{center}
\par \vspace{2mm}
\begin{center}
{\bf S. Catani}\\

\vspace{5mm}

{I.N.F.N., Sezione di Firenze}\\
{and Dipartimento
di Fisica, Universit\`a di Firenze}\\
{Largo E. Fermi 2, I-50125 Florence, Italy}

\vspace{5mm}

{\bf M.H. Seymour}\\

\vspace{5mm}

{Theory Division, CERN}\\
{CH-1211 Geneva 23, Switzerland}
\end{center}

\par \vspace{2mm}
\begin{center} {\large \bf Abstract} \end{center}
\begin{quote}
  In order to make quantitative predictions for jet cross sections in
  perturbative QCD, it is essential to calculate them to next-to-leading
  accuracy. This has traditionally been an extremely laborious process. Using
  a new formalism, imaginatively called the dipole formalism, we are able to
  construct a completely general algorithm for next-to-leading order
  calculations of arbitrary jet quantities in arbitrary processes. In this
  paper we present the basic ideas behind the algorithm and illustrate them
  with a simple example.
\end{quote}
\vspace*{\fill}
\begin{flushleft}
     CERN-TH/96-28 \\   February 1996
\end{flushleft}
\end{titlepage}

\renewcommand{\thefootnote}{\fnsymbol{footnote}}

\section{Introduction}
\label{int}

\vspace*{-2mm}
Most of the recent progress in the understanding of strong interaction physics
at large momentum transfer has been due to the comparison between precise 
experimental data and very accurate QCD calculations to higher perturbative 
orders [\ref{QCDrev}].

These higher-order computations have been carried out over a period of
about fifteen years, often long after the accuracy of experimental data
has made them necessary, because of the difficulties in setting up a
general and straightforward calculational procedure. The physical origin of 
these difficulties is in the necessity of factorizing the long- and 
short-distance components of the scattering processes
and is reflected in the perturbative 
calculation by the presence of divergences.

In general, when evaluating higher-order QCD cross sections,
one has to consider 
real-emission contributions and virtual-loop corrections and one has to deal
with different kind of singularities. The customary {\em ultraviolet\/}
singularities, present in the virtual contributions, are removed by
renormalization. The low-momentum ({\em soft\/}) and small-angle 
({\em collinear\/}) regions instead produce singularities both in the real and 
in the virtual contributions. In order to handle these divergences, the 
observable one is interested in has to be properly defined. It has to be a 
{\em jet quantity}, that is, a hadronic observable that turns out to be 
infrared safe and either collinear safe or collinear factorizable: its actual 
value 
has to be independent of the number of soft and collinear particles in the 
final state (see Sect.~\ref{siee} for a formal definition). 
In the case of jet 
quantities, the coherent sum over different (real and virtual) soft and 
collinear partonic configurations in the final state leads to the cancellation 
of soft singularities. The left-over collinear singularities are then 
factorized into the process independent structure and fragmentation functions 
of partons (parton distributions), leading to predictable scaling violations. 
As a result, jet cross sections are finite (calculable) at the partonic level 
order by order in perturbation theory. All the dependence on long-distance 
physics is either included in the parton distributions or in non-perturbative 
corrections that are suppressed by inverse powers of the (large)
transferred momentum $Q$ that controls the scattering process.

Because of this pattern of singularities, 
QCD calculations of jet cross sections beyond leading order (LO)  
are very
involved. Owing to the complicated phase space for multi-parton 
configurations, analytic calculations are in practice impossible for 
all but the simplest quantities, but the use of numerical methods is 
far from trivial because soft and collinear singularities present in the
intermediate steps have first to be regularized. This is usually done by
analytic continuation to a number of space-time dimensions $d=4-2\epsilon$
different from four, which greatly complicates the Lorentz algebra in the
evaluation of the matrix elements and prevents a straightforward implementation
of numerical 
integration techniques. Despite these difficulties, efficient computational
techniques have been set up, at least to next-to-leading order (NLO), 
during the last few years. 

There are, broadly speaking, two types of algorithm used for NLO
calculations: one based on the phase-space {\em slicing\/} method and the 
other
based on the {\em subtraction\/} method\footnote{We refer the reader to the 
Introduction of Ref.~[\ref{KS}] for an elementary description of the basic 
difference between the two methods.}. The main difference between these 
algorithms and the standard procedures of analytic calculations is that only 
a minimal part of the full calculation is treated analytically, namely only 
those contributions giving rise to the singularities. Moreover, for any given
process, these contributions are computed in a manner that is independent of
the particular jet observable considered. Once every singular term has been
isolated and the cancellation/factorization of divergences achieved, one can 
perform the remaining part of the calculation in four space-time dimensions.
Although, when possible, one still has the freedom of completing 
the calculation analytically, at this point the use of numerical integration 
techniques (typically, Monte Carlo methods) is certainly more convenient. 
First of all, the numerical approach allows one to calculate any number and 
any type of observable simultaneously by simply histogramming the appropriate
quantities, rather than having to make a separate analytic calculation for each
observable. Furthermore, using the numerical approach, it is easy to 
implement different experimental conditions, for example, detector acceptances
and experimental cuts. In other words, the phase-space slicing and subtraction
algorithms provide the basis for setting up a general-purpose Monte 
Carlo program for carrying out arbitrary NLO QCD calculations in a given
process.
 
Both the slicing [\ref{KL}] and the 
subtraction [\ref{ERT}] methods were first used in the context of NLO 
calculations of three-jet cross sections in \ee\ annihilation. Then they
have been applied to other cross sections adapting the method each time
to the particular process. Only recently has it
become clear that both algorithms are generalizable in a process-independent 
manner. The key observation is that the singular parts of the QCD matrix
elements for real emission can be singled out in a general way by using
the factorization properties of soft and collinear radiation [\ref{BCM}].

At present, a general slicing algorithm is available for
calculating NLO cross sections for {\em any\/} number of jets both in
lepton [\ref{GG}] and hadron [\ref{GGK}] collisions. To our knowledge, 
fragmentation processes have been considered only in the particular case of 
direct-photon production [\ref{BOO}]. 
The complete generalization of this method to include fragmentation functions 
and heavy flavours is still missing.

As for the subtraction algorithm, a general NLO formalism has been set up 
for computing three-jet observables in \ee\ annihilation [\ref{ERT},\ref{KN}]
and cross sections up to two final-state jets [\ref{KS},\ref{EKS}]
in hadron collisions\footnote{An extension of the
method for three-jet cross sections has been recently presented
[\ref{Frix}].}. Also the treatment of massive partons has been 
considered in the particular case of heavy-quark correlations in hadron 
collisions [\ref{MNR}].

In this paper, we present the basic idea to set up a {\em completely general\/}
version of the  subtraction algorithm. This generality is obtained by fully
exploiting the factorization properties of soft and collinear emission and, 
thus, deriving new improved factorization formulae, called {\em dipole 
factorization formulae}. They allow us to introduce a set of universal 
counter-terms that can be used for {\em any\/} NLO QCD calculation. 

For the purpose of illustration, in this short contribution we describe the
implementation of our general method to the calculation of jet cross sections
in processes with no initial-state hadrons, typically \ee\ annihilation.
Full details of the method and its application to all the other hard-scattering
processes will appear elsewhere [\ref{CS}]. 

We first recall the main features of the subtraction method in
Section~\ref{gen}. Then in Section~\ref{dff} we present our dipole
factorization formulae, for the case in which there are no incoming QCD
partons. These allow us to calculate cross-sections for an arbitrary number of
jets in \ee\ annihilation, which we do in Section~\ref{siee}.  After briefly
recapping the resulting formulae in Section~\ref{finfor}, we illustrate them
\enlargethispage*{1mm}
with a simple example in Section~\ref{3jet}, $e^+e^-\to3$ jets.  Finally in
Section~\ref{outl} we give a summary and outlook.

\section{The subtraction procedure}
\label{gen}

Suppose we want to compute a jet cross section $\sigma$ to NLO, namely
\beq 
\label{sig}
\sigma = \sigma^{LO} + \sigma^{NLO} \;.
\eeq
Here the LO cross section $\sigma^{LO}$ is obtained by integrating the 
exclusive cross section $d\sigma^{B}$ in the Born approximation over the phase
space for the corresponding jet quantity. Suppose also that this LO calculation 
involves $m$ partons in the final state. Thus, we write
\beq
\label{sLO} 
\sigma^{LO} = \int_m d\sigma^{B} \;,
\eeq
where, in general, all the quantities (QCD matrix elements and 
phase space) are evaluated in $d=4-2\ep$ space-time dimensions. However, by 
definition, at this LO the phase space integration in Eq.~(\ref{sLO}) is finite
so that the whole calculation can be carried out (analytically or numerically)
in four dimensions.

Now we go to NLO\@. We have to consider the exclusive cross section
$d\sigma^{R}$
with $m+1$ partons in the final-state and the one-loop correction $d\sigma^{V}$
to the process with $m$ partons in the final state:
\beq
\label{sNLO}
\sigma^{NLO} \equiv \int d\sigma^{NLO} =
 \int_{m+1} d\sigma^{R} + \int_{m} d\sigma^{V} \;.
\eeq

The two integrals on the right-hand side of Eq.~(\ref{sNLO}) are separately
divergent if $d=4$, although their sum is finite. Therefore, before any 
numerical calculation can be attempted, the separate pieces have to be 
regularized. Using dimensional regularization, the divergences (arising out
of the integration) are replaced
by double (soft and collinear) poles $1/\ep^2$ and single (soft, collinear
or ultraviolet) poles $1/\ep$. Suppose that one has already carried out the
renormalization procedure in $d\sigma^{V}$ so that all its ultraviolet poles
have been removed.

The general idea of the subtraction method for writing a general-purpose 
Monte Carlo program is to use the identity
\beq
\label{dsNLO}
d\sigma^{NLO} = \left[ d\sigma^{R} - d\sigma^{A}  \right]
+  d\sigma^{A} +  d\sigma^{V} \;\;,
\eeq
where $d\sigma^{A}$ is a proper approximation of $d\sigma^{R}$ such as to have
the same {\em pointwise\/} singular behaviour (in $d$ dimensions) as 
$d\sigma^{R}$ itself.
Thus, $d\sigma^{A}$ acts as a {\em local\/} counterterm for 
$d\sigma^{R}$ and, introducing the phase space integration,
\beq
\label{sNLO1}
\sigma^{NLO} = \int_{m+1} \left[ d\sigma^{R} - d\sigma^{A}  \right]
+  \int_{m+1} d\sigma^{A} +  \int_m d\sigma^{V} \;\;,
\eeq
one can safely perform the limit $\ep \to 0$ under the integral sign in the 
first
term on the right-hand side of Eq.~(\ref{sNLO1}). Hence, this first term can be
integrated numerically in four dimensions.

All the singularities are now associated to the last two terms on the
right-hand side of Eq.~(\ref{sNLO1}). If one is able to carry out analytically
the integration of $d\sigma^{A}$ over the one-parton subspace leading to the
$\ep$ poles, one can combine these poles with those in $d\sigma^{V}$, thus 
cancelling all the divergences, performing the limit $\ep \to 0$ and 
carrying out numerically the remaining integration over the $m$-parton phase 
space. The final structure of the calculation is as follows
\beq
\label{sNLO2}
\sigma^{NLO} =
\int_{m+1} \left[ d\sigma^{R}_{\ep=0} - d\sigma^{A}_{\ep=0} \right]
+  \int_m \left[ d\sigma^{V} +  \int_1 d\sigma^{A} \right]_{\ep=0} \;\;,
\eeq
and can be easily implemented in a \lq partonic Monte Carlo' program, which
generates appropriately weighted partonic events with $m+1$ final-state partons
and events with $m$ partons.

The key for the subtraction procedure to work is obviously 
the actual form of $d\sigma^{A}$. 
One needs to find an expression for $d\sigma^{A}$ that fulfils the following
properties: $i)$~for any given process, $d\sigma^{A}$ has to be obtained in a 
way
that is independent of the particular jet observable considered; $ii)$~it has 
to exactly match the singular behaviour of $d\sigma^{R}$ in $d$ dimensions;
$iii)$~its form has to be 
particularly convenient for Monte Carlo integration techniques; 
$iv)$~it has to be exactly integrable analytically in $d$ 
dimensions over the single-parton subspaces leading to soft and collinear 
divergences.

In Ref.~[\ref{ERT}], a suitable expression for $d\sigma^{A}$ for the process
\ee$\to 3$~jets was obtained by starting from the explicit expression (in
$d$ dimensions) of the corresponding $d\sigma^{R}$ and by performing extensive
partial fractioning of the $3+1$-parton matrix elements, so that each divergent
piece could be extracted. This is an extremely laborious and ungeneralizable
task, in the sense that having done it for \ee$\to 3$~jets does not help
us to do this for, say, \ee$\to 4$~jets or for any other process.

In Ref.~[\ref{KS}], the general properties of soft and collinear emission were
first used (in the context of the subtraction method) to construct 
$d\sigma^{A}$, for one- and two-jet production in hadron collisions, in a way 
that is independent of the detailed form of the corresponding $d\sigma^{R}$.

The central proposal of our version of the subtraction method is that one can
give a recipe for constructing $d\sigma^{A}$ that is completely {\em process 
independent\/} (and not simply independent of the jet observable). 
Starting from our physical knowledge of how the $m+1$-parton matrix elements
behave in the soft and collinear limits that produce the divergences, we 
introduce improved factorization formulae, called
dipole formulae (see Sect.~\ref{dff}), which
allow us to obtain in a straightforward way (see Sect.~\ref{siee}) a 
counter-term $d\sigma^A$ 
satisfying all the properties listed above.

\section{Dipole factorization formulae}
\label{dff}

\noindent {\it Notation}
\vspace{0.1cm}

In general we use dimensional regularization in $d=4-2\ep$
space-time dimensions and consider $d-2$ helicity states for gluons
and 2 helicity states for massless quarks. This defines the usual 
dimensional-regularization scheme. Other dimensional-regularization
prescriptions can be used, as well [\ref{CS}].

The dimensional-regularization scale, which appears in the calculation
of the matrix elements, is denoted by $\mu$. 
In the perturbative calculation of physical cross sections, 
after having combined the renormalized matrix elements, 
the dependence on $\mu$ exactly cancels and is replaced by the dependence on 
the renormalization scale $\mu_R$. Therefore, in order to avoid a cumbersome 
notation, we set $\mu=\mu_R$. 

Throughout the paper, $\as$ stands for $\as(\mu)$,
the NLO QCD running coupling evaluated at the renormalization scale $\mu$.
The actual value of the QCD coupling $\as(\mu)$ depends on the 
renormalization scheme used to subtract the ultraviolet divergences
from the (bare) one-loop matrix element (or, equivalently, from $d\sigma^V$
in Eq.~(\ref{sNLO})). 

The $d$-dimensional phase space, which involves the integration over the 
momenta $\{p_1, ...,$ $p_m\}$ of $m$ final-state partons, will be denoted as 
follows
\beq
\label{psm}
\left[ \;\prod_{l=1}^{m} \frac{d^dp_l}{(2\pi)^{d-1}}
\,\delta_+(p_l^2) \right] \;(2\pi)^{d} \,\delta^{(d)}(p_1 + ... + p_m - Q)
\equiv d\phi_m(p_1, ...,p_m;Q) \;\;.
\eeq

In the case of processes without initial-state
QCD partons (\ee-type processes),
the ({\em tree-level\/}) matrix element with $m$ QCD partons in the final state
has the following general structure (non-QCD partons, namely 
$\gamma^*, Z^0, W^{\pm}, \ldots$, carrying
a total incoming momentum $Q_\mu$, are always understood)
\beq
\cm_m^{c_1, ...,c_m; s_1, ...,s_m}(p_1, ...,p_m)
\eeq
where $\{c_1, ...,c_m\}$, $\{s_1, ...,s_m\}$ and 
$\{p_1,..,p_m\}$ are respectively colour indices
($a=1, ...,$ \mbox{$N_c^2-1$} different colours for each gluon,  
$\alpha=1, ..,N_c$ different colours for each quark or antiquark),
spin indices ($\mu=1,...,d$ for gluons, $s=1,2$ for massless fermions)
and momenta.

It is useful to introduce a basis 
$\{ \ket{c_1,...,c_m} \otimes \ket{s_1,...,s_m} \}$
in colour + helicity space in such a way that
\beq
\cm_m^{c_1, ...,c_m; s_1, ...,s_m}(p_1, ...,p_m) \equiv
\Bigl( \bra{c_1,...,c_m} \otimes \bra{s_1,...,s_m} \Bigr) \mket{1, ..., m} \;.
\eeq
Thus $\mket{1, ..., m}$ is a vector in colour + helicity space.
According to this notation, the matrix element squared (summed
over final-state colours and spins) $\M{}{m}$ can be written as
\beq
\M{}{m} = {}_m\!\!<{1, ..., m}|{1, ..., m}>_m \;.
\eeq

As for the colour structure\footnote{Within our formalism, there is no need
to consider the decomposition of the matrix elements into colour subamplitudes,
as in [\ref{GG},\ref{MP}].}, it is convenient to associate a colour charge
${\bf T}_i$ with the emission of a gluon from each parton $i$. We thus define
the square of colour-correlated tree-amplitudes as follows
\beeq
\label{colam}
|{\cal{M}}_{m}^{i,k}|^2 &\equiv& 
{}_m\!\!<{1, ..., m}| \,{\bom T}_i \cdot {\bom T}_k \,|{1, ..., m}>_m 
\nonumber \\
&=&
\left[ {\cal M}_m^{a_1.. b_i ... b_k ... a_m}(p_1,...,p_m) \right]^*
\; T_{b_ia_i}^c \, T_{b_ka_k}^c 
\; {\cal M}_m^{a_1.. a_i ... a_k ... a_m}(p_1,...,p_m) \;,
\eeeq
where $T_{c b}^a \equiv i f_{cab}$ (colour-charge matrix
in the adjoint representation)  if the emitting particle $i$ 
is a gluon and $T_{\alpha \beta}^a \equiv t^a_{\alpha \beta}$
(colour-charge matrix in the fundamental representation) 
if the emitting particle $i$ is a quark (in the case of an emitting
antiquark $T_{\alpha \beta}^a \equiv {\bar t}^a_{\alpha \beta}
= - t^a_{\beta \alpha }$). It is straightforward to check that 
the colour-charge algebra is:
\beq
{\bom T}_i \cdot {\bom T}_j ={\bom T}_j \cdot {\bom T}_i \;\;\;\;{\rm if}
\;\;i \neq j; \;\;\;\;\;\;{\bom T}_i^2= C_i,
\eeq
where $C_i$ is the Casimir operator, that is,
$C_i=C_A=N_c$ if $i$ is a gluon and $C_i=C_F=(N_c^2-1)/2N_c$ if $i$ is a quark
or antiquark.

In this notation, each vector $\mket{1, ..., m}$ is a colour singlet, so
colour conservation is simply
\beq
\sum_{i=1}^m {\bom T}_i \mket{1, ..., m} = 0.
\eeq

\vspace{0.3cm}
\noindent {\it Dipole formulae}
\vspace{0.1cm}

The real contribution $d\sigma^R$ to the NLO cross section in Eq.~(\ref{sNLO})
is proportional to the tree-level matrix element ${\cal M}_{m+1}$ for 
producing $m+1$ partons in the final state.
The dependence of $|\cm_{m+1}|^2$
on the momentum $p_j$ of a final-state parton $j$ is singular in two different
phase-space regions: when the momentum $p_j$ vanishes ({\em soft\/} 
region) and/or when it becomes parallel to the momentum $p_i$ of
another parton in $\cm_{m+1}$ ({\em collinear\/} region).
This singular behaviour of $|\cm_{m+1}|^2$ is well-known [\ref{BCM},\ref{AP}] 
and universal.
Indeed, in the soft and 
collinear limits, $\cm_{m+1}$ is essentially factorizable 
with respect to $\cm_{m}$, the tree-level amplitude with $m$ partons,
and the singular factor only depends
on the momenta and quantum numbers of the QCD partons in $\cm_{m}$.

We thus introduce the following  dipole factorization formula: 
\beq \label{ff}
\!\!|{\cal M}_{m+1}(p_1, ...,p_{m+1})|^2 = 
\obra{1, ..., m+1} \oket{1, ..., m+1} = \sum_{k \neq i,j}
{\cal D}_{ij,k}(p_1, ...,p_{m+1}) + \dots
\eeq
where $\dots$ stands for terms that are not singular in the
limit $p_i \cdot p_j \rightarrow 0$ (i.e.\ when $i$ and $j$ become collinear or
when either $i$ or $j$ is soft) and the dipole contribution
${\cal D}_{ij,k}$ is given by
\beeq \label{dipff}
\!\!\!&{\cal D}_{ij,k}&\!\!\!\!(p_1, ...,p_{m+1}) =
\frac{-1}{2 p_i \cdot p_j}  \\
&&\cdot \;\; 
\mbra{1, .., {\widetilde {ij}},.., {\widetilde k},.., m+1} 
\,\frac{{\bom T}_k \cdot {\bom T}_{ij}}{{\bom T}_{ij}^2} \; {\bom V}_{ij,k} \,
\mket{1, .., {\widetilde {ij}},.., {\widetilde k},.., m+1} \;.
\nonumber 
\eeeq

The $m$-parton matrix element on the right-hand side of Eq.~(\ref{dipff}) is
obtained from the original $m+1$-parton matrix element by replacing $a)$ the
partons $i$ and $j$ with a single parton ${\widetilde {ij}}$ (which plays the 
role of {\em emitter\/})
and $b)$ the parton $k$ with the parton $\widetilde k$ 
(which plays the role of {\em spectator\/}).
All of the quantum numbers (colour, flavour) except momenta are assigned as 
follows. The spectator
parton $\widetilde k$ has the same quantum numbers as $k$. The quantum numbers
of the emitter parton ${\widetilde {ij}}$ are obtained according to their
conservation in the collinear splitting process ${\widetilde {ij}} \to i + j$
(i.e.\ \lq anything + gluon' gives \lq anything' and \lq quark + antiquark'
gives \lq gluon').

The momenta of the emitter and the spectator are defined as follows
\beq \label{pk}
{\widetilde p}_{ij}^\mu = 
p_i^\mu  + p_j^\mu  - \frac{y_{ij,k}}{1-y_{ij,k}} \,p_k^\mu \;\;, 
\;\;\;\;{\widetilde p}_k^\mu = \frac{1}{1-y_{ij,k}} \,p_k^\mu \;\;, 
\eeq
where the dimensionless variable $y_{ij,k}$ is given by
\beq \label{yijk}
y_{ij,k} = \frac{p_ip_j}{p_ip_j+p_jp_k+p_kp_i} \;.
\eeq

In the bra-ket on the right-hand side of Eq.~(\ref{dipff}), 
${\bom T}_{ij}$ and ${\bom T}_k$ are the colour charges of the emitter
and the spectator and ${\bom V}_{ij,k}$ are matrices in the helicity space 
of the emitter. These matrices, which depend on $y_{ij,k}$ and
on the kinematic variables ${\tilde z}_i, {\tilde z}_j$:
\beeq
\label{zitil}
{\tilde z}_i 
= \frac{p_ip_k}{p_jp_k+p_ip_k} =
\frac{p_i{\widetilde p}_k}{{\widetilde p}_{ij}{\widetilde p}_k}
\;, \;\;\;  
{\tilde z}_j = \frac{p_jp_k}{p_jp_k+p_ip_k}= 
\frac{p_j{\widetilde p}_k}{{\widetilde p}_{ij}{\widetilde p}_k}=
1-{\tilde z}_i  \;,
\eeeq  
are universal factors related to the $d$-dimensional Altarelli-Parisi 
splitting functions [\ref{AP}].
For fermion + gluon splitting we have
($s$ and $s'$ are the spin indices of the fermion ${\widetilde {ij}}$
in $\bra{..,{\widetilde {ij}},..}$ and $\ket{..,{\widetilde {ij}},..}$
respectively)
\beeq
\label{vqgk}
\bra{s}
{\bom V}_{q_ig_j,k}({\tilde z}_i;y_{ij,k}) \ket{s'}
&=& 8\pi \mu^{2\ep} \as\; C_F\,
\left[ \frac{2}{1-{\tilde z}_i(1-y_{ij,k})} 
- \frac{}{} (1+{\tilde z}_i)
-\ep (1-{\tilde z}_i)\right] \;\delta_{ss'} \nonumber \\ 
&\equiv&  V_{q_ig_j,k} \;\delta_{ss'} \;. 
\eeeq
For quark + antiquark and gluon + gluon splitting we have
($\mu$ and $\nu$ are the spin indices of the gluon ${\widetilde {ij}}$
in $\bra{..,{\widetilde {ij}},..}$ and $\ket{..,{\widetilde {ij}},..}$
respectively)
\beq \label{vqbqk}
\bra{\mu}
{\bom V}_{q_i{\bar q}_j,k}({\tilde z}_i)
\ket{\nu}
= 8\pi \mu^{2\epsilon} \as \;
T_R \;
\left[ -g^{\mu \nu} - \frac{2}{p_ip_j} \, 
( {\tilde z}_i p_i^{\mu} - {\tilde z}_j p_j^{\mu} )
\,( {\tilde z}_i p_i^{\nu} - {\tilde z}_j p_j^{\nu} ) \,\right] 
\equiv  V_{q_i{\bar q}_j,k}^{\mu \nu} \,,
\eeq
\beeq \label{vggk}
&\!&\!\!\!\!\!\!\!\!\!\!\bra{\mu}{\bom V}_{g_ig_j,k}({\tilde z}_i;y_{ij,k})
\ket{\nu} = 16\pi 
\mu^{2\epsilon} \as \;C_A\,
\left[ -g^{\mu \nu} \left( 
\frac{1}{1-{\tilde z}_i(1-y_{ij,k})} \right. \right.\nonumber \\
&\!&\!\!\!\!\!\!\!\!\!\!+ \left. \left.
\frac{1}{1-{\tilde z}_j(1-y_{ij,k})} - 2 \right)  
+ (1-\ep) \frac{1}{p_ip_j} \, 
( {\tilde z}_i p_i^{\mu} - {\tilde z}_j p_j^{\mu} )
\,( {\tilde z}_i p_i^{\nu} - {\tilde z}_j p_j^{\nu} ) \,\right] 
\equiv  V_{g_ig_j,k}^{\mu \nu} \;.
\eeeq

The factorization formula in Eq.~(\ref{ff}) has a dipole structure with respect
to the {\em colour\/} and {\em spin\/} indices of the factorized partons. 
As shown in Ref.~[\ref{CS}], Eqs.~(\ref{ff},\ref{dipff}) coincide with
the soft-gluon [\ref{BCM}] and Altarelli-Parisi [\ref{AP}] factorization
formulae respectively in the soft and collinear limits. However, 
Eqs.~(\ref{ff},\ref{dipff}) are completely well-defined also outside these
limiting regions of the phase space. Indeed, in the factorized $m$-parton 
matrix element both the emitter ${\widetilde {ij}}$ and the spectator
${\widetilde k}$ are on-shell 
$({\widetilde p}_{ij}^2 = {\widetilde p}_k^2 = 0)$ and, performing the 
replacement $\{i,j,k\} \to \{{\widetilde {ij}},{\widetilde k}\}$, momentum 
conservation is implemented exactly:
\beq \label{momcon}
p_i^\mu  + p_j^\mu + p_k^\mu = {\widetilde p}_{ij}^\mu +
{\widetilde p}_k^\mu \;\;.
\eeq
The importance of these kinematical features is twofold. Firstly, momentum
conservation leads to a smooth interpolation between the soft and collinear
limits, thus avoiding double counting of overlapping soft and collinear
singularities. Secondly, the definition (\ref{pk}) of the dipole
momenta allows us to factorize exactly the $m+1$-parton phase space into an
$m$-parton subspace times a single-parton contribution (see 
Eqs.~(\ref{psfac},\ref{dpi})). 
The first property allows us to construct a counter-term $d\sigma^A$ that
produces a pointwise cancellation of the singularities of $d\sigma^R$ as in
Eq.~(\ref{sNLO1}). The second property makes this counter-term fully integrable
analytically over the subspace leading to soft and collinear divergences.

\section{The calculation of jet cross sections}
\label{siee}

The dipole formulae form the basis for our general algorithm for NLO jet
calculations, as we describe below.  First we define the leading order cross
section and give a formal definition of the requirements a jet definition must
fulfil.  Then we introduce the subtraction term, which cancels all
singularities of the real matrix element, and show how it can be integrated in
$d$ dimensions to cancel the singularities of the virtual matrix element.

\vspace{0.3cm}
\noindent {\it Leading order and jet definition}
\vspace{0.1cm}

The Born-level cross section in Eq.~(\ref{sLO}) has the following expression
\beq
\label{dsmb}
d\sigma^{B} = {\cal N}_{in} \sum_{\{ m \} } \, d\phi_m(p_1, ...,p_m;Q)
\;\frac{1}{S_{\{ m \} }} \;|{\cal{M}}_{m}(p_1, ...,p_m) |^2
\;F_J^{(m)}(p_1, ...,p_m) \;\;,
\eeq
where ${\cal N}_{in}$ includes all the factors that are QCD independent,
$\sum_{\{ m \} }$ denotes the sum over all the configurations with $m$
partons, $d\phi_m$ is the partonic phase space in Eq.~(\ref{psm}),
$S_{\{ m \} }$ is the Bose symmetry factor for identical partons in the final 
state and
${\cal{M}}_{m}$ is the tree-level matrix element.

The function $F_J^{(m)}(p_1, ...,p_m)$ defines the jet observable in terms of 
the momenta of the 
$m$ final-state partons. In general, $F_J$ may contain $\theta$-functions
(thus, Eq.~(\ref{dsmb}) defines precisely a cross section), $\delta$-functions
(Eq.~(\ref{dsmb}) defines a differential cross section), numerical and 
kinematical factors (Eq.~(\ref{dsmb}) refers to an inclusive observable), or
any combination of these. The essential property of $F_J^{(m)}$ is that the
jet observable we are interested in has to be infrared
and collinear safe. That is, it has to be 
experimentally (theoretically) defined in such a 
way that its actual value is independent of the number of soft and 
collinear hadrons (partons) produced in the final state. In particular, this
value has to be the same in a given $m$-parton configuration and in all
$m+1$-parton configurations that are kinematically degenerate with it
(i.e.\ which are obtained from the $m$-parton configuration by adding a soft 
parton or replacing a parton with a pair of collinear partons carrying the same
total momentum). These properties can be simply restated in a formal way, as
follows
\beq
\label{fjsoft}
F_J^{(n+1)}(p_1,..,p_j,..,p_{n+1} ) \to 
F_J^{(n)}(p_1,...,p_{n+1} )  \;\;\;\;\;\ {\rm if} \;\;p_j \to 0 \;\;,
\eeq
\beq
\label{fjcoll}
F_J^{(n+1)}(p_1,..,p_i,..,p_j,..,p_{n+1} ) \to
F_J^{(n)}(p_1,..,p_i+p_j,..,p_{n+1} ) \;\;\;\;\;\ 
{\rm if} \;\;p_i \parallel p_j \;,
\eeq
\beq
\label{fjLO}
F_J^{(m)}(p_1,...,p_m ) \to 0 \;\;\;\;\;\ {\rm if} \;\;p_i \cdot p_j 
\to 0 \;\;.
\eeq
Equations (\ref{fjsoft}) and (\ref{fjcoll}) respectively guarantee that the
jet observable is infrared and collinear safe for {\em any\/} number $n$ of
final-state partons, i.e.\ to {\em any\/} order in QCD perturbation theory.
Equation (\ref{fjLO}) defines the LO cross section, that is, it 
ensures that the Born-level cross section $d\sigma^{B}$ in Eq.~(\ref{dsmb})
is well-defined (i.e.\ finite after integration) in $d=4$ dimensions.

\vspace{0.3cm}
\noindent {\it Next-to-leading order: the subtraction term} 
\vspace{0.1cm}

The real contribution $d\sigma^{R}$ to the NLO cross section in 
Eq.~(\ref{sNLO})
has the same expression as $d\sigma^{B}$ in Eq.~(\ref{dsmb}), 
apart from the replacement $m \to m + 1$.
In particular, the $m$-parton matrix element $\cm_m$ is replaced by
$\cm_{m+1}$. Therefore 
an explicit and general form for the local counter-term $d\sigma^{A}$ in
Eq.~(\ref{dsNLO}) is provided
by the dipole factorization formula (\ref{ff}):
\beeq
\label{dsadef}
d\sigma^{A} &=& {\cal N}_{in} \sum_{\{ m+1 \} } \, 
d\phi_{m+1}(p_1, ...,p_{m+1};Q)
\; \frac{1}{S_{\{ m+1 \} }} \nonumber \\
&\cdot& \sum_{\mathrm{pairs}\atop i,j}
\;\sum_{k\not=i,j} {\cal D}_{ij,k}(p_1, ...,p_{m+1}) \;
F_J^{(m)}(p_1, .. {\widetilde p}_{ij}, {\widetilde p}_k, ..,p_{m+1}) \;\;.
\eeeq
Here ${\cal D}_{ij,k}(p_1, ...,p_{m+1})$ is the dipole contribution 
in Eq.~(\ref{dipff})
and $F_J^{(m)}(p_1, .. {\widetilde p}_{ij}, {\widetilde p}_k, ..,p_{m+1})$
is the jet function for the corresponding $m$-parton state
$\{ p_1, .. {\widetilde p}_{ij}, {\widetilde p}_k, ..,p_{m+1} \}$.
Note that this is completely independent of $p_i,$ which is how we are able to
integrate $d\sigma^A$ analytically over the phase-space of $i$ without any
information about the form of $F_J,$ as we perform below.

We can check that the definition (\ref{dsadef}) makes the difference
$( d\sigma^{R} - d\sigma^{A} )$ integrable in $d=4$ dimensions. Its 
explicit expression is
\beeq
\label{dsra}
d\sigma^{R} - d\sigma^{A} &=& {\cal N}_{in} \sum_{\{ m+1 \} } \, 
d\phi_{m+1}(p_1, ...,p_{m+1};Q)
\; \frac{1}{S_{\{ m+1 \} }}  \nonumber \\
&\cdot& \left\{ \frac{}{} |{\cal{M}}_{m+1}(p_1, ...,p_{m+1}) |^2
\;F_J^{(m+1)}(p_1, ...,p_{m+1}) \frac{}{} \right. \nonumber \\
&-& \left.
\sum_{\mathrm{pairs}\atop i,j}
\;\sum_{k\not=i,j} \right. \left. \frac{}{}
 {\cal D}_{ij,k}(p_1, ...,p_{m+1}) \;
F_J^{(m)}(p_1, .. {\widetilde p}_{ij}, {\widetilde p}_k, ..,p_{m+1}) \right\}
\;\;.
\eeeq
Each term in the curly bracket is separately singular in the soft and 
collinear regions. However, as stated in Sect.~\ref{dff}, in each of these 
regions both the matrix element ${\cal{M}}_{m+1}$ and the phase space
for the $m+1$-parton configuration behave as the corresponding dipole 
contribution and dipole phase space:
\beeq
|{\cal{M}}_{m+1}(p_1, ...,p_{m+1}) |^2 &\to& {\cal D}_{ij,k}(p_1, ...,p_{m+1})
\;\;, \\
\{ p_1, .. p_i,..p_j,..p_k, ..,p_{m+1} \} &\to& 
\{ p_1, .. {\widetilde p}_{ij}, {\widetilde p}_k, ..,p_{m+1} \} \;\;.
\eeeq
Thus, because of Eqs.~(\ref{fjsoft}) and (\ref{fjcoll}), the singularities
of the first term in the curly bracket are cancelled by similar singularities
due to the second term. On the other hand, each dipole ${\cal D}_{ij,k}$ in 
Eq.~(\ref{dipff}) has no other singularities but those due to the $m$-parton 
matrix element $\mket{1, .., {\widetilde {ij}},.., {\widetilde k},.., m+1}$.
Because of Eq.~(\ref{fjLO}), these singularities are screened (regularized)  
by the jet 
function $F_J^{(m)}(p_1, .. {\widetilde p}_{ij}, {\widetilde p}_k, ..,p_{m+1})$
in the curly bracket of Eq.~(\ref{dsra}).

Note that this cancellation mechanism is completely independent of the actual
form of the jet defining function but it is essential that $d\sigma^R$ and
$d\sigma^A$ are proportional to $F_J^{(m+1)}$ and $F_J^{(m)}$ respectively.
Nonetheless, because the terms on the 
left- and  right-hand sides of Eq.(\ref{ff}) depend on the same kinematic
variables, $\{ p_1, ...,p_{m+1} \},$
both $d\sigma^R$ and $d\sigma^A$ live on the same $m+1$-parton
phase space. Thus the numerical integration (in $d=4$ 
dimensions) of Eq.~(\ref{dsra}) via Monte Carlo techniques is straightforward. 
One
simply generates an $m+1$-parton configuration and gives it a positive
($+ \,|{\cal{M}}_{m+1}|^2$) or negative ($- \,{\cal D}_{ij,k}$) weight. 
The role
of the two different jet functions $F_J^{(m+1)}$ and $F_J^{(m)}$
is that of binning these weighted events into different bins of the jet 
observable. Any time that the generated configuration approaches a singular
region, these two bins coincide and the cancellation of the large  positive 
and negative weights takes place.

Note, also, that the helicity dependence
of the splitting kernels ${\bom V}_{ij,k}$ in Eqs.~(\ref{vqbqk},\ref{vggk})
is essential if $d\sigma^A$ is to act as a local counterterm  that makes 
$[ d\sigma^R - d\sigma^A ]$ integrable in four dimensions.  Indeed, the parton 
azimuthal correlations due to this dependence are not only essential in the
most general case when $F_J$ explicitly depends on them, but even when it does
not\footnote{In this case the evaluation of 
$\int_{m+1} d\sigma^R$ in four dimensions usually involves double angular
integrals of the type $\int_{-1}^{+1} d\cos\theta \int_{0}^{2\pi} d\varphi
\;\cos\varphi/(1-\cos\theta),$ where $\varphi$ is the azimuthal angle. These  
integrals are mathematically ill-defined. If their numerical integration is
attempted, one can obtain any answer whatsoever, depending on the detail of the
integration procedure.  Performing the integral analytically before going to 4
dimensions, one obtains $\int_{-1}^{+1} d\cos\theta \int_{0}^{2\pi} d\varphi
\;\cos\varphi/(1-\cos\theta)\sin^{-2\ep}\theta\sin^{-2\ep}\varphi=0$.}.

\vspace{0.3cm}
\noindent {\it Next-to-leading order: integral of the subtraction term} 
\vspace{0.1cm}

Having discussed the four-dimensional integrability of 
$( d\sigma^{R} - d\sigma^{A} )$, the only other step we have to consider
is the $d$-dimensional analytical integrability of $d\sigma^{A}$ over the
one-parton subspace leading to soft and collinear divergences.
In this respect, our dipole formalism is particularly efficient and simple.
The definition (\ref{pk}) of the dipole momenta 
allows us to {\em exactly\/} factorize the phase space of the partons $i,j,k$
into the dipole phase space times a single-parton contribution, as follows 
\beq \label{psfac}
d\phi_{m+1}(p_1,..,p_i,p_j,p_k,..,p_{m+1};Q) =
d\phi_m(p_1,..,{\widetilde p}_{ij},
{\widetilde p}_k, ..,p_{m+1};Q) \;\left[ dp_i({\widetilde p}_{ij},
{\widetilde p}_k) \right] \;\;,
\eeq
where
\beeq
\label{dpi}
\left[ dp_i({\widetilde p}_{ij},{\widetilde p}_k) \right]
= \frac{d^{d}p_i}{(2\pi)^{d-1}} \,\delta_+(p_i^2) \; \Theta(1- {\tilde z}_i) 
\,\Theta(1-y_{ij,k}) \;
\frac{(1-y_{ij,k})^{d-3}}{1- {\tilde z}_i} \;,
\eeeq
and the kinematic variables $y_{ij,k}$ and ${\tilde z}_i$ are defined in
Eqs.~(\ref{yijk},\ref{zitil}).

Inserting Eq.~(\ref{psfac}) and the explicit expression (\ref{dipff}) for
${\cal D}_{ij,k}$ into Eq.~(\ref{dsadef}), we can factorize completely the 
$p_i$ dependence and carry out the integration over the phase space region
(\ref{dpi}) with the $m$ parton momenta 
$\{ p_1,..,{\widetilde p}_{ij},{\widetilde p}_k, ..,p_{m+1} \}$ kept fixed.
Remarkably, this integration can be exactly performed in closed analytical form
in any number of space-time dimensions.
As shown in detail in Ref.~[\ref{CS}], after integration the spin
correlations between ${\bom V}_{ij,k}$ and 
$\mket{1, .., {\widetilde {ij}},.., {\widetilde k},.., m+1}$ vanish and only 
colour correlations survive.
The final result for $\int_{m+1} d\sigma^A$ can be written
in terms of an $m$-parton integral of the LO 
(colour-correlated) matrix element times a factor [\ref{CS}]:
\beeq
\label{intdsma}
\int_{m+1} d\sigma^{A} = \int_m \left[ \int_1 d\sigma^A \right] &=&
\int_m {\cal N}_{in} \sum_{\{ m \} } \, d\phi_m(p_1, ...,p_m;Q)
\;\frac{1}{S_{\{ m \} }} \nonumber \\
&\cdot& 
{}_m\!\!<{1, ..., m}| \;{\bom I}(\ep) \;|{1, ..., m}>_m \;
F_J^{(m)}(p_1, ...,p_m) \;\;.
\eeeq
Comparing Eqs.~(\ref{intdsma}) and (\ref{dsmb}), we see that the integration
of $d\sigma^A$ over the one-parton subspace that produces soft and collinear
singularities leads to an expression completely analogous to $d\sigma^B$.
One should simply replace the matrix element squared
$\M{}{m} =$ \mbox{${}_m\!\!<{1, ..., m}|{1, ..., m}>_m$} in $d\sigma^B$ 
with
\beq
\label{insop}
{}_m\!\!<{1, ..., m}| \;{\bom I}(\ep) \;|{1, ..., m}>_m \;, 
\eeq
where ${\bom I}(\ep)$ is an insertion operator that depends on the colour 
charges
and momenta of the $m$ final-state partons.
Its explicit expression is [\ref{CS}]:
\beeq
\label{iee}
{\bom I}(p_1,...,p_m;\ep) = - 
\frac{\as}{2\pi}
\frac{1}{\Gamma(1-\ep)} \sum_i \;\frac{1}{{\bom T}_{i}^2} \;{\cal V}_i(\ep)
\; \sum_{k \neq i} {\bom T}_i \cdot {\bom T}_k
\; \left( \frac{4\pi \mu^2}{2 p_i\cdot p_k} \right)^{\ep}
 \;\;,
\eeeq
where the singular factors ${\cal V}_i(\ep)$ 
have the following $\ep$-expansion\footnote{Their exact expressions in 
any number $d=4-2\ep$ of dimensions are given in [\ref{CS}].} 
\beeq
\label{calvexp}
{\cal V}_{i}(\ep) = {\bom T}_{i}^2 \left( \frac{1}{\ep^2} - 
\frac{\pi^2}{3} \right) + \gamma_i \;\frac{1}{\ep}
+ \gamma_i + K_i + {\cal O}(\ep) \;\;,
\eeeq
with ($T_R=1/2$ and $N_f$ is the number of flavours)
\beq
\label{gadef}
\gamma_{i=q,{\bar q}} = \mbox{$\frac{3}{2}$}\, C_F \;,
\;\;\;\;
\gamma_{i=g} = \mbox{$\frac{11}{6}$} \, C_A - \mbox{$\frac{2}{3}$}
T_R N_f \;,
\eeq
\beeq
\label{kcondef}
K_{i=q,{\bar q}} = \left( \mbox{$\frac{7}{2}$} - \mbox{$\frac{\pi^2}{6}$}
\right) C_F \;,
\;\;\;\;
K_{i=g} = \left( \mbox{$\frac{67}{18}$} - \mbox{$\frac{\pi^2}{6}$} \right) C_A
- \mbox{$\frac{10}{9}$} T_R N_f \;.
\eeeq

In the calculation of the NLO cross section (\ref{sNLO2}), Eq.~(\ref{intdsma})
has to be combined with the virtual contribution, whose expression  
in terms of the
(renormalized) one-loop matrix element is the following
\beq
\label{dsv1}
d\sigma^{V} = {\cal N}_{in} \sum_{\{ m \} } \, d\phi_m(p_1, ...,p_m;Q)
\;\frac{1}{S_{\{ m \} }} \;|{\cal{M}}_{m}(p_1, ...,p_m) |^2_{(1-loop)}
\;F_J^{(m)}(p_1, ...,p_m) \;\;.
\eeq
As discussed in Ref.~[\ref{CS}], the addition of these two contributions
correctly produces the cancellation of all the $\ep$-poles, thus leading 
to a finite NLO cross section.

\section{Final formulae}
\label{finfor}

The final results of the application of our algorithm to the calculation
of jet cross sections with no hadron in the initial state are summarized
below.

The full QCD cross section in Eq.~(\ref{sig}) contains a LO and a NLO 
component.
Assuming that the LO calculation involves $m$ final-state partons, the LO
cross section is given by
\beq
\label{LOeefin}
\sigma^{LO} = \int_m d\sigma^{B} = \int d\Phi^{(m)} 
\;| \cm_{m}(p_1, ...,p_m)|^2 \;F_J^{(m)}(p_1, ...,p_m) \;\;,
\eeq
where $\cm_{m}$ is the tree-level QCD matrix element for producing $m$ partons
in the final state and the function $F_J^{(m)}$ defines the particular jet 
observable we are interested in (see Eqs.~(\ref{fjsoft}-\ref{fjLO}) 
for the general 
properties that $F_J^{(m)}$ has to fulfil). The factor $d\Phi^{(m)}$ collects
all the relevant phase space factors, i.e.\ all the remaining terms
on the right-hand side of Eq.~(\ref{dsmb}). The whole calculation (phase space
integration and evaluation of the matrix element) can be carried out in 
four space-time dimensions.

According to the subtraction formula, Eq.~(\ref{sNLO2}), 
the NLO cross section
is split into two terms with $m+1$-parton and  $m$-parton kinematics, 
respectively. The contribution with $m+1$-parton kinematics is the following
\beeq
\label{m1eefin}
\int_{m+1} \left[ d\sigma^R_{\ep=0} - d\sigma^A_{\ep=0} \right]&=&
\int d\Phi^{(m+1)}  \left\{ \frac{}{}
\;| \cm_{m+1}(p_1, ...,p_{m+1})|^2 \;F_J^{(m+1)}(p_1, ...,p_{m+1}) \right. \\
&&\left. - \sum_{\mathrm{pairs}\atop i,j}
\;\sum_{k\not=i,j} {\cal D}_{ij,k}(p_1, ...,p_{m+1}) \;
F_J^{(m)}(p_1, .. {\widetilde p}_{ij}, {\widetilde p}_k, ..,p_{m+1}) \right.
\left. \frac{}{} \right\}
\;\;, \nonumber
\eeeq 
where the term in the curly bracket is exactly the same as that in 
Eq.~(\ref{dsra}): $\cm_{m+1}$ is the tree-level matrix element, 
${\cal D}_{ij,k}$ is the dipole factor in Eq.~(\ref{dipff}) and $F_J^{(m)}$ 
is the
jet defining function for the corresponding $m$-parton state (note, again, 
the difference between the two jet functions $F_J^{(m+1)}$ and $F_J^{(m)}$
in the curly bracket). In spite of their original $d$-dimensional
definition, at this stage the full calculation is carried out in four 
dimensions.

The NLO contribution with $m$-parton kinematics is given by
\beeq
\label{meefin}
&&\!\!\!\!\!\!\!\int_{m} \left[ d\sigma^{V}  + \int_1 d\sigma^{A} 
 \right]_{\ep=0}  \\
&&\!\!\!\!\!\!\! 
=  \int d\Phi^{(m)} \left\{ | \cm_{m}(p_1, ...,p_{m})|^2_{(1-loop)}
+ {}_m\!\!<{1, ..., m}| \;{\bom I}(\ep) \;|{1, ..., m}>_m \right\}_{\ep=0}
 F_J^{(m)}(p_1, ...,p_{m}) \;\;.\nonumber
\eeeq
The first term in the curly bracket is the one-loop {\em renormalized\/} matrix
element squared for producing $m$ final-state partons. The second term is 
obtained by inserting the colour-charge operator of Eq.~(\ref{iee}) into the 
tree-level matrix element for producing $m$ partons as in Eq.~(\ref{insop}).
These two terms have to be first evaluated in $d=4-2\ep$ 
dimensions. Then one has to carry out their expansion in $\ep$-poles (the 
expansion for the singular factors ${\cal V}_i(\ep)$ is given in
Eq.~(\ref{calvexp})),
cancel analytically (by trivial addition) the poles and perform the limit 
$\ep \to 0$. At this point the phase-space integration is carried out in four
space-time dimensions.

\section{\boldmath\ee$\to 3$ jets} 
\label{3jet}

In this Section we consider the simplest non-trivial application of our 
algorithm, namely the calculation of three-jet observables in \ee\ annihilation.
Thus our formalism can be directly compared with that in Ref.~[\ref{ERT}].

The LO partonic process to be considered is 
$e^+ e^- \to q(p_1) + {\bar q}(p_2)  + g (p_3)$. The corresponding tree-level
matrix element is denoted by $\cm_3(p_1,p_2,p_3)$. We use customary
notation for the kinematic variables:
$Q^2$ is the square of the centre-of-mass energy, $y_{ij}=2p_i\cdot p_j/Q^2$
and $x_i=2p_i\cdot Q/Q^2$.

At NLO, two different real-emission subprocesses contribute:  
a) $e^+ e^- \to q(p_1) + {\bar q}(p_2)  + g (p_3) + g(p_4)$;
b) $e^+ e^- \to q(p_1) + {\bar q}(p_2)  + q(p_3) + {\bar q}(p_4)$. In addition
one has to compute the one-loop correction to the LO process.

The calculation of the subtracted cross section (\ref{m1eefin}) for the 
subprocess
a) involves the evaluation of the following dipole contributions:
${\cal D}_{13,2}, {\cal D}_{13,4}, {\cal D}_{14,2}, {\cal D}_{14,3},
{\cal D}_{23,1}, {\cal D}_{23,4},
{\cal D}_{24,1},$ ${\cal D}_{24,3}, {\cal D}_{34,1}, {\cal D}_{34,2}$.
The associated colour
algebra can be easily performed in closed form because the several colour
projections of the three-parton matrix element fully 
factorize\footnote{If the LO matrix element involves two or three partons,
the colour algebra can always be carried out in closed (factorized) form.}.
Thus we do not need to compute any colour-correlated tree amplitudes
and we find
\beeq
\label{dipa}
{\cal D}_{13,2}(p_1,p_2,p_3,p_4) &=& \frac{1}{2p_1p_3} 
\left(1- \frac{C_A}{2C_F}\right) 
V_{q_1g_3,2} \;|\cm_3({\widetilde p_{13}},{\widetilde p_2},p_4)|^2 \;\;,
\nonumber \\
{\cal D}_{13,4}(p_1,p_2,p_3,p_4) &=& \frac{1}{2p_1p_3} \,\frac{C_A}{2C_F} \; 
V_{q_1g_3,4} \;|\cm_3({\widetilde p_{13}},p_2,{\widetilde p_4})|^2 \;\;, \\
{\cal D}_{34,1}(p_1,p_2,p_3,p_4) &=& \frac{1}{2p_3p_4} \,\frac{1}{2} \; 
V_{g_3g_4,1}^{\mu \nu} 
\;{\cal T}_{\mu \nu}({\widetilde p_{1}},p_2,{\widetilde p_{34}}) \;\;.
\nonumber
\eeeq
The dipole contributions ${\cal D}_{23,1}, {\cal D}_{23,4}, 
{\cal D}_{34,2}$ are obtained respectively 
from ${\cal D}_{13,2},{\cal D}_{13,4}, {\cal D}_{34,1} $ by the replacement 
$p_1 \leftrightarrow p_2$. Analogously, one can obtain 
${\cal D}_{14,2}$ and ${\cal D}_{14,3}$ respectively from
${\cal D}_{13,2}$ and ${\cal D}_{13,4}$ by the replacement
$p_3 \leftrightarrow p_4$, and ${\cal D}_{24,1}$ and 
${\cal D}_{24,3}$ respectively from ${\cal D}_{13,2}$ and
${\cal D}_{13,4}$ by the replacement $p_1 \leftrightarrow p_2, \;
p_3 \leftrightarrow p_4$.

In the case of the subprocess b) we have to consider the following
dipole contributions:
${\cal D}_{12,3}, {\cal D}_{12,4}, {\cal D}_{14,2}, {\cal D}_{14,3}, 
{\cal D}_{23,1}, {\cal D}_{23,4}, {\cal D}_{34,1}, {\cal D}_{34,2}$.
Performing the colour algebra we get
\beq
\label{dipb}
{\cal D}_{34,1}(p_1,p_2,p_3,p_4) = \frac{1}{2p_3p_4} \frac{1}{2} \; 
V_{q_3{\bar q}_4,1}^{\mu \nu} 
{\cal T}_{\mu \nu}({\widetilde p_{1}},p_2,{\widetilde p_{34}}) \;,
\eeq
and all the other dipoles are obtained by the corresponding permutation of 
the parton momenta.

The splitting functions $V_{ij,k}$ of Eqs.~(\ref{dipa},\ref{dipb}) are given 
in Eqs.~(\ref{vqgk}-\ref{vggk}). The tensor ${\cal T}_{\mu \nu}$ is 
the squared amplitude for the LO process $\mathrm{e^+e^-\to q\bar{q}g}$
not summed over the gluon polarizations ($\mu$ and $\nu$ are the gluon spin
indices and $-g^{\mu \nu} {\cal T}_{\mu \nu} = |\cm_3|^2 $).  
This can be easily calculated. In the case of jet observables averaged over
the directions of the incoming leptons (un-oriented events) we find
(in $d=4$ dimensions)
\beq
{\cal T}^{\mu \nu}(p_1,p_2,p_3) = - \frac{1}{x_1^2 + x_2^2}   
\; |\cm_3(p_1,p_2,p_3)|^2 \;T^{\mu \nu} \;,
\eeq
where
\beeq
  T^{\mu\nu} &=&
    +2                            \frac{p_1^\mu p_2^\nu}{Q^2}
    +2                            \frac{p_2^\mu p_1^\nu}{Q^2}
    -2\frac{1-x_1}{1-x_2}         \frac{p_1^\mu p_1^\nu}{Q^2}
    -2\frac{1-x_2}{1-x_1}         \frac{p_2^\mu p_2^\nu}{Q^2}
\nonumber\\&&
    -\frac{1-x_1-x_2+x_2^2}{1-x_2}\left[
                                  \frac{p_1^\mu   p_3^\nu}{Q^2}
    +                             \frac{p_3^\mu p_1^\nu}{Q^2}
                                  \right]
    -\frac{1-x_2-x_1+x_1^2}{1-x_1}\left[
                                  \frac{p_2^\mu p_3^\nu}{Q^2}
    +                             \frac{p_3^\mu p_2^\nu}{Q^2}
                                  \right]
\nonumber\\&&
    +\left(1+\hf x_1^2+\hf x_2^2-x_1-x_2\right)g^{\mu\nu} \;.
\eeeq

To complete the NLO calculation we also need the virtual cross section. 
In the case of un-oriented events, we take the one-loop matrix element
in the ${\overline {\rm MS}}$ renormalization scheme from Ref.~[\ref{ERT}]
(we use slightly different notation):
\beeq
\label{ert1loop}
&&|\cm_3(p_1,p_2,p_3)|^2_{(1-loop)} =
|\cm_3(p_1,p_2,p_3)|^2 \; \frac{\as}{2\pi}\frac{1}{\Gamma(1-\epsilon)}   
\left(\frac{4\pi\mu^2}{Q^2}\right)^\epsilon
\nonumber \\
&&\cdot 
\left\{ \,-\frac{1}{\ep^2} \left[ (2C_F -C_A) y_{12}^{-\ep} + 
C_A \left(y_{13}^{-\ep} + y_{23}^{-\ep} \right) \right]
-\frac{1}{\ep} \left( 3C_F + \frac{11}{6}C_A -\frac{2}{3}T_R N_f \right)
\right. \nonumber \\
&&+ \left. \frac{\pi^2}{2}(2C_F+C_A) - 8 C_F \right\} 
+ \frac{\as}{2\pi}\left[ F(y_{12},y_{13},y_{23}) +{\cal O}(\ep) \right] \;\;,
\eeeq
where $F(y_{12},y_{13},y_{23})$ is defined in Eq.~(2.21) of Ref.~[\ref{ERT}].

The explicit evaluation of the insertion operator ${\bom I}(\ep)$ in
Eqs.~(\ref{insop},\ref{iee}) gives: 
\beeq
\label{ertinsop}
&&{}_3\!\!<{1,2,3}| \;{\bom I}(\ep) \;|{1,2,3}>_3 = 
|\cm_3(p_1,p_2,p_3)|^2 \; \frac{\as}{2\pi}\frac{1}{\Gamma(1-\epsilon)}
\left(\frac{4\pi\mu^2}{Q^2}\right)^\epsilon   
\nonumber \\
&&
\left\{ \,\frac{1}{\ep^2} \left[ (2C_F -C_A) y_{12}^{-\ep} +
C_A \left(y_{13}^{-\ep} + y_{23}^{-\ep} \right) \right]
+\frac{1}{\ep} \left( 2\gamma_q + \gamma_g \right) \right.
\nonumber \\
&&- \left. \gamma_q \frac{1}{C_F} \left[ (2C_F - C_A) \ln y_{12} + \frac{1}{2}
C_A \ln (y_{13}y_{23}) \right] - \frac{1}{2} \gamma_g \ln (y_{13}y_{23})
\right. \nonumber \\
&&\left. -\frac{\pi^2}{3}(2C_F+C_A) + 2(\gamma_q + K_q) + \gamma_g + K_g  
+{\cal O}(\ep) \right\} \;\;.
\eeeq
Combining the one-loop  matrix element (\ref{ert1loop}) with the result 
(\ref{ertinsop}) according to Eq.~(\ref{meefin}) and using the explicit 
expressions (\ref{gadef},\ref{kcondef}) for $\gamma_i$ and $K_i$,
all the pole terms cancel. 
Note that as well as the pole terms, the closely related $\pi^2$ and
$\ln^2$ terms cancel:
\beeq
&&|\cm_3(p_1,p_2,p_3)|^2_{(1-loop)} + 
{}_3\!\!<{1,2,3}| \;{\bom I}(\ep) \;|{1,2,3}>_3 = |\cm_3(p_1,p_2,p_3)|^2
\nonumber \\
&&\cdot \;\frac{\as}{2\pi} 
\left [ - \frac{3}{2} (2C_F - C_A) \ln y_{12}
-\frac{1}{3} (5C_A - T_R N_f) \ln (y_{13}y_{23}) \right. \nonumber \\
&&\left. + 2 C_F +\frac{50}{9} C_A
- \frac{16}{9} T_R N_f \right] 
+ \frac{\as}{2\pi}\left[ F(y_{12},y_{13},y_{23}) +{\cal O}(\ep) \right]] \;\;.
\eeeq 

\begin{figure}
  \centerline{\epsfig{figure=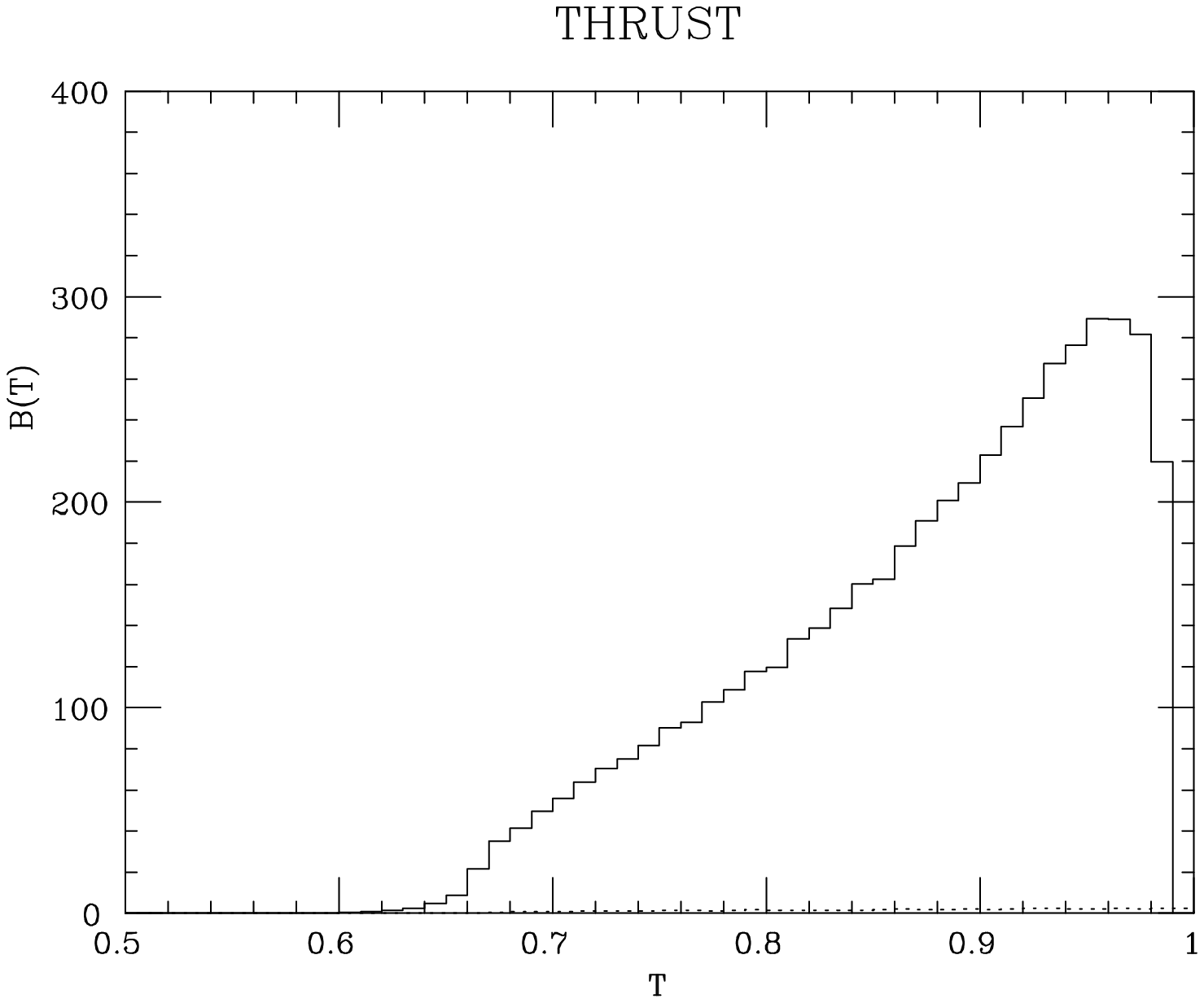,height=6cm}\hfill
              \epsfig{figure=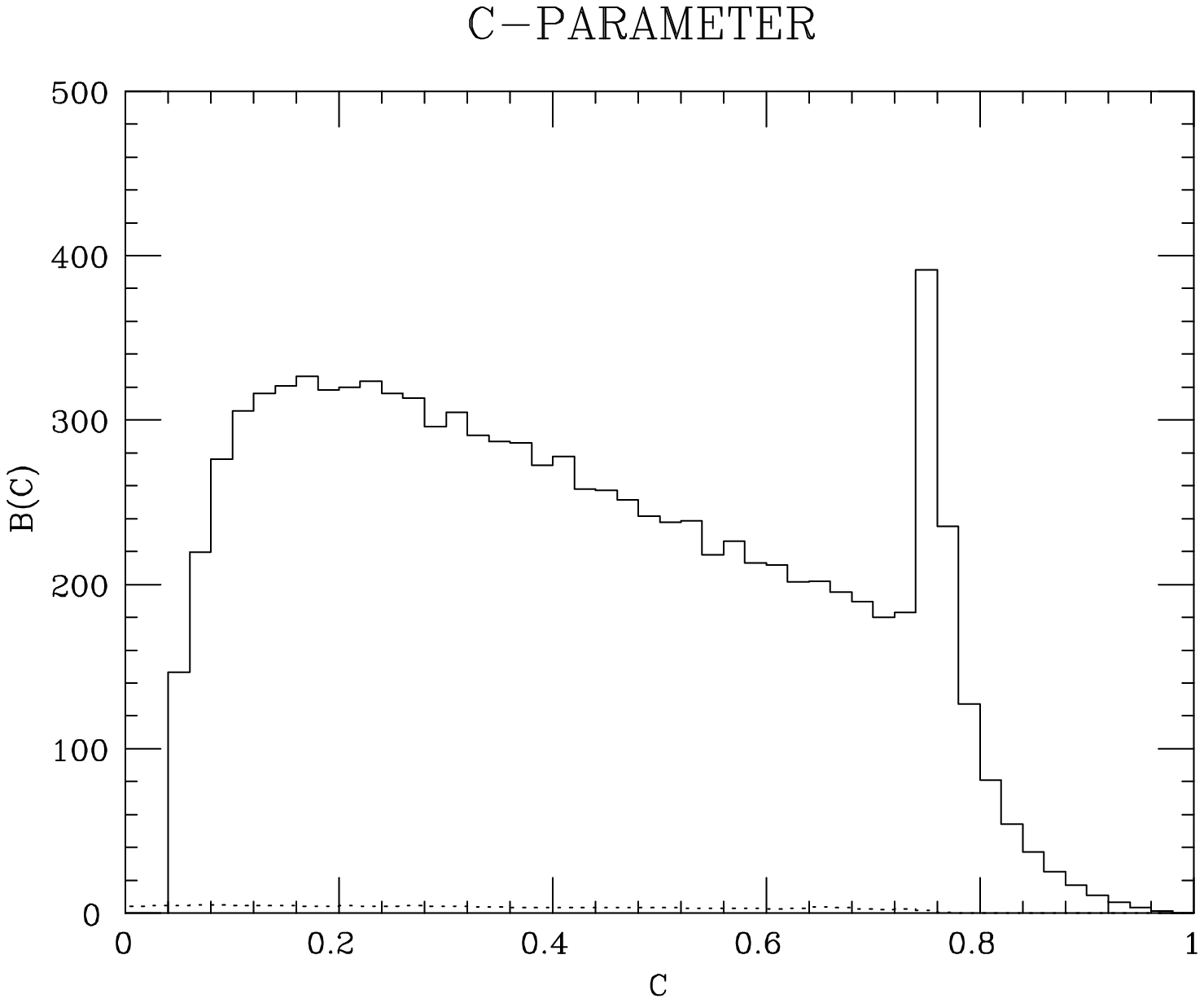,height=6cm}}
  \caption[]{Coefficient of $(\as/2\pi)^2$ for the thrust and $C$-parameter
    distributions. The dotted histograms show the size of the statistical
    errors.}
  \label{thefig}
\end{figure}

We have implemented these results as a working Monte Carlo program\footnote{The
  program can be obtained from
  \verb+http://surya11.cern.ch/users/seymour/nlo/+.} and the results are in
good agreement with Ref.~[\ref{KN}] for all distributions shown there.  As an
example we show the NLO coefficients for the thrust and $C$-parameter
distributions in Fig.~\ref{thefig}.
We find that in general, the numerical convergence is similar to the program of
Ref.~[\ref{KN}], except close to the two-jet region in which ours becomes
progressively better. More details of the Monte Carlo program and the
generalization to oriented three-jet events will be presented elsewhere.

\section{Summary and outlook}
\label{outl}

In this letter we have presented the basic idea to set up a completely 
general algorithm for calculating jet cross sections in NLO QCD\@. By general 
we mean that the algorithm applies to {\em any\/} jet observable in a given
scattering process as well as to {\em any\/} hard-scattering process. The
algorithm overcomes all the analytical difficulties related to the treatment
of soft and collinear divergences in the perturbative expansion. The output
of the algorithm is given in terms of effective matrix elements (the 
contributions in the curly bracket of Eqs.~(\ref{m1eefin},\ref{meefin})) 
with built-in
cancellation of soft and collinear singularities. These effective matrix 
elements can be numerically or analytically (whenever possible) integrated
over the available phase space to compute the actual value of the NLO cross 
section. If the numerical approach is chosen, Monte Carlo integration 
techniques can be easily implemented to provide a general-purpose Monte
Carlo program for carrying out NLO QCD calculations in any given process.

Starting from the very general idea of the subtraction method, we have
discussed how one can automatically construct a pointwise and integrable
counter-term $d\sigma^A$ for the real contribution $d\sigma^R$ to the  NLO
cross section. This counter-term is independent of the details of the process
under study and can be handled once and for all to define the effective 
matrix elements mentioned above. We have shown how this procedure works in
practice for processes with no hadron in the initial state (typically,
\ee\ annihilation) and any number of jets in the final state. In this case, the
final result is given by the master formulae in 
Eqs.~(\ref{m1eefin},\ref{meefin}). We
have provided explicit expressions for both the universal dipole factors 
${\cal D}_{ij,k}$ and $\bom I$. Having  
these factors at our disposal, the only other ingredients necessary for the 
full NLO calculation, are the following
\begin{itemize}
\addtolength{\itemsep}{-4pt}
\item
a set
of independent colour projections\footnote{Actually, if the total number of 
QCD partons involved in the LO matrix element is less than or equal to three,
one simply needs its incoherent sum over the colours (see Sect.~\ref{3jet}).} 
of the matrix element 
squared at the Born level, summed over parton polarizations, in $d$
dimensions;
\item
the one-loop matrix element in $d$ dimensions;
\item
an additional projection of the Born level matrix element over the helicity of
each external gluon in four dimensions;
\item
the tree-level NLO matrix element in four dimensions.
\end{itemize}

\noindent These few ingredients are sufficient for writing, in a 
straightforward way,
a general-purpose NLO Monte Carlo algorithm. Note in particular
that there is no need to extract a proper counter-term $d\sigma^{A}$ 
starting from a 
cumbersome expression for $d\sigma^{R}$ in $d$ dimensions. The NLO matrix
element 
contributing to $d\sigma^{R}$ can be evaluated directly in four space-time 
dimensions thus leading to an extreme simplification of the Lorentz algebra.

The key point of our method for constructing the counter-term $d\sigma^A$ 
is the dipole formalism described in Sect.~\ref{dff}. Starting from the 
universal
behaviour of the QCD matrix elements in the soft and collinear regions, we have
introduced improved factorization formulae based on a dipole structure with
respect to the momenta, colours and helicities of the QCD partons. Our dipole
formulae correctly match the well-known singularities of the QCD scattering 
amplitudes in the soft and collinear limits. Moreover, these limits are
approached smoothly, thus avoiding double counting of overlapping soft and
collinear divergences. This smooth transition is possible because our dipole 
formalism is explicitly Lorentz covariant and the dipole formulae exactly
fulfil momentum conservation.

In the present paper, these main features of the dipole formulae have been used
in the context of NLO computations of jet cross sections with no initial-state
hadron. For lepton-hadron and hadron-hadron collisions, perturbative
QCD calculations face additional difficulties related to the factorization of
initial-state collinear singularities. Likewise, to calculate cross sections
for processes in which a final state hadron is identified, the equivalent
final-state collinear singularities must be dealt with. In a companion paper
[\ref{CS}], we show that the dipole formalism overcomes these difficulties in
a simple and general manner. Thus we can explicitly provide a set of universal
counter-terms or, equivalently, effective (non-singular) matrix elements
that can be straightforwardly used for {\em any\/} NLO QCD calculation.  

\subsection*{Acknowledgements}
One of us (SC) would like to thank the CERN TH Division for hospitality and
partial support during the course of this work.

\subsection*{References} 
\begin{enumerate}

\item \label{QCDrev}
S.\ Catani, in Proc.\ of the Int.\ Europhysics Conf.\ on High Energy Physics,
HEP 93, 
eds.\ J.\ Carr and M.\ Perrottet (Editions
Frontieres, Gif-sur-Yvette, 1994), p.~771; B.R.\ Webber, 
in Proc.\ of the 27th Int.\ Conf.\ on High Energy Physics,
eds. P.J.\ Bussey and I.G.\ Knowles (Institute of Physics,
Philadelphia, 1995), p.~213.

\item \label{KS}
Z.\ Kunszt and D.E.\ Soper, \pr{46}{192}{92}.

\item \label{KL}
K.\ Fabricius, G.\ Kramer, G.\ Schierholz and I.\ Schmitt, \zp{11}{315}{81};
G.\ Kramer and B.\ Lampe, Fortschr. Phys. 37 (1989) 161.

\item \label{ERT}
R.K.\ Ellis, D.A.\ Ross and A.E.\ Terrano,
\np{178}{421}{81}.

\item  \label{BCM}
See, for instance: A.\ Bassetto, M.\ Ciafaloni and G.\ Marchesini,
\prep{100}{201}{83}; Yu.L.\ Dokshitzer, V.A.\ Khoze, A.H.\ Mueller and
S.I.\ Troyan, {\em Basics of Perturbative QCD}, Editions Frontieres,
Paris, 1991.

\item \label{GG}
W.T.\ Giele and E.W.N.\ Glover, \pr{46}{1980}{92}.
 
\item \label{GGK}
W.T.\ Giele, E.W.N.\ Glover and D.A.\ Kosower, \np{403}{633}{93}.

\item \label{BOO}
H.\ Baer, J.\ Ohnemus and J.F.\ Owens, \pr{42}{61}{90}; B.\ Bailey,
J.F.\ Owens and J.\ Ohnemus, \pr{46}{2018}{92}.

\item  \label{KN}
Z.\ Kunszt and P.\ Nason, in \lq Z Physics at LEP 1', CERN 89-08, vol.~1,
p.~373.

\item \label{EKS}
S.D.\ Ellis, Z.\ Kunszt and D.E.\ Soper, \pr{40}{2188}{89}, \prl{69}{1496}{92}.

\item \label{Frix}
S.\ Frixione, Z.\ Kunszt and A.\ Singer, preprint SLAC-PUB-95-7073, 
hep-ph/9512328.

\item \label{MNR}
M.L.\ Manga\-no, P.\ Nason and G.\ Ridolfi, \np{373}{295}{92}.

\item \label{CS}
S.\ Catani and M.H.\ Seymour, preprint CERN-TH/96-29, to appear.

\item \label{MP}
M.L.\ Mangano and S.J.\ Parke, \prep{200}{301}{91}.

\item \label{AP}
G.\ Altarelli and G.\ Parisi, \np{126}{298}{77}.

\end{enumerate}

\end{document}